\documentclass[aps,prd,preprint,floats,tightenlines]{revtex4}

\newcommand{\beq}{\begin{equation}}
\newcommand{\eeq}{\end{equation}}

\begin{document}

\title{Freak observers and the measure of the multiverse}

\author{Alexander Vilenkin}

\address{
Institute of Cosmology, Department of Physics and Astronomy,\\ 
Tufts University, Medford, MA 02155, USA
}

\begin{abstract}

I suggest that the factor $p_j$ in the pocket-based measure of the
multiverse, $P_j=p_j f_j$, should be interpreted as accounting for
equilibrium de Sitter vacuum fluctuations, while the selection factor
$f_j$ accounts for the number of observers that were formed due to
non-equilibrium processes resulting from such fluctuations. I show
that this formulation does not suffer from the problem of freak
observers (also known as Boltzmann brains). 

\end{abstract}

\maketitle

\section{Introduction}

The simplest interpretation of the observed accelerated expansion of
the universe is that it is driven by a constant vacuum energy density,
$\rho_v=const$, which is about 3 times greater than the density of
nonrelativistic matter. Ordinary matter is being diluted, while the
vacuum energy density remains the same, and in another 10 billion
years or so the universe will be compltely dominated by the
vacuum. The following evolution of the universe is accurately
described by de Sitter space.

It has been shown by Gibbons and Hawking \cite{GH} that the state of
quantum fields in de Sitter space is similar to a thermal state with a
characteristic temperature $T_{GH}=H/2\pi$, where \beq H=(8\pi
G\rho_v/3)^{1/2} \eeq is the de Sitter expansion rate. For the
observed value of $\rho_v$, the Gibbons-Hawking temperature is
extremely low, $T_{GH}\sim 10^{-29}$~K.  Nevertheless, interesting
things will occasionally pop out of the vacuum as quantum
fluctuations, at a nonzero rate per unit spacetime volume. An
intelligent observer, like a human, could be one such thing. Or, short
of a complete observer, a disembodied brain may fluctuate into
existence, with a pattern of neuron firings creating a perception of
being on Earth and observing the CMB radiation.  Of course, the
nucleation rate $\Gamma_F$ of such freak observers (also known as
Boltzmann brains \cite{Rees,Rees1,Albrecht}) is extremely small
\cite{DKS02,Page1,Page2}. But the important point is that it is
nonzero.

De Sitter space is eternal to the future, so no matter how small
$\Gamma_F$ is, freak observers will eventually outnumber regular
observers who have ever lived in the universe
\cite{Susskind,Page06,BF06}.  Regular observers are formed as a result of
non-equilibrium processes which started at the big bang and will
eventually end when the universe thermalizes at the temperature
$T_{GH}$.  The total number of such observers that will exist in a
fixed comoving volume is finite. On the other hand, the cumulative
number of freak observers grows unboundedly with time. (In fact, it
grows exponentially, since the corresponding physical volume grows as
$\exp(3Ht)$.)  Then the question is: Why are we not freak observers?
(Assuming that we believe we are not.)

This issue has been recently discussed by Page \cite{Page06}, who
concluded that the least unattractive way for us to avoid being freaks
is to require that our vacuum should be rather unstable and should
decay within a few Hubble times of the vacuum domination, that is, in
20 billion years or so.

Before accepting such a drastic conclusion, we need to analyze the
situation in some more detail. Two important facts that need to be
taken into account are (i) that our local universe appears to be a
product of cosmic inflation and (ii) that inflation is generically
eternal. Then bubbles of high-energy false vacuum can nucleate in our
low-energy vacuum \cite{LeeWeinberg}. Such bubbles become sites of
eternal inflation, producing an infinite number of pocket universes
like ours, each containing an infinite number of observers
\cite{recycling}.  The nucleation rate of false vacuum bubbles may be
much lower that that of freak observers. But considering that each
bubble nucleation yields an infinite number of regular obsevers, one
might conclude that regulars totally outnumber the freaks
\cite{Lindeprivate}.

The trouble is that in an eternally inflating universe the numbers of
both types of observers are infinite. They can be meaningfully
compared only if one adopts some prescription to regulate the
infinities. A related issue, which has recently attracted much
attention, is the calculation of probabilities for different vacua in
multiverse models, also known as the measure problem. A number of
prescriptions have been proposed, and one can try to apply them to the
problem of freak observers.  A ``holographic'' measure \cite{Bousso},
which restrict consideration to the part of the universe within the
causal diamond of a single observer, can resolve the problem if our
vacuum is sufficiently unstable, enough for the vacuum decay to
prevent freak domination \cite{BF06}. Some ways to avoid the problem
using a measure based on a globally defined time coordinate have been
discussed in \cite{Linde06,Page3}.

Here, I am going to address the problem of freak observers in the context
of the pocket-based measure, which was introduced in \cite{GSPVW} and
which satisfies the physically reasonable requirements of
gauge-invariance and independence of initial conditions (see also
\cite{AV06}). In a recent paper, Bousso and Freivogel argued that this
measure predicts freak domination and should therefore be ruled
out. In fact, the formulation of the measure prescription in
\cite{GSPVW} disregarded the existence of freaks and was therefore
incomplete. With freaks taken into account, the prescription as it
stands gives meaningless infinite answers for the probabilities. I am
going to suggest how the problem can be fixed by clarifying the
formulation of the pocket-based measure.

\section{The pocket-based measure}

The pocket-based prescription for the measure is a two-step
procedure. The probability $P_j$ for a randomly picked observer to be
in a pocket of type $j$ is given by the product
\beq
P_j=p_j f_j,
\label{Pj}
\eeq
where $p_j$ is an abundance of bubbles (pockets) of type $j$ and $f_j$
is the selection factor characterizing the relative number of
observers in different pockets.  To calculate the bubble abundance
$p_j$, one first chooses a future-directed congruence of geodesics and
a segment of a spacelike hypersurface $\Sigma$ which is crossed by
that congruence.  The geodesics project bubbles in the future of
$\Sigma$ back onto $\Sigma$, and the prescription of \cite{GSPVW} is
to find what fraction of all bubbles is of type $j$, counting only
bubbles whose projected size is greater than $\epsilon$, and then take
the limit $\epsilon\to\infty$,
\beq
p_j=\lim_{\epsilon\to 0}{N_j(>\epsilon)\over{N(>\epsilon)}}.
\eeq
The resulting $p_j$ are independent of the choice of the geodesic
congruence and of the hypersurface $\Sigma$. The bubble count is
dominated by the bubbles nucleating in the asymptotic future, so the
result is independent of the initial conditions at the onset of
inflation.  (An equivalent prescription for $p_j$ has been suggested
in \cite{Easther}.)

The prescription for the selection factor $f_j$ is that it is given by
the total number of independent observers that evolve in a fixed
comoving volume,
\beq
f_j \propto R^3 \int_{\tau_{in}}^\infty n_j(\tau)a_j^3(\tau)d\tau.
\label{fj}
\eeq
Here, $R$ is a fixed comoving length scale, the same for all bubbles,
$n_j(\tau)$ is the average number of observers produced in a pocket
$j$ per unit physical volume per unit time and $a_j(\tau)$ is the
scale factor in that pocket. The time coordinate $\tau$ is the proper
time in the standard open FRW coordinates inside the bubble. The
initial time $\tau_{in}$ is arbitrary, as long as it is chosen small
enough. (At small $\tau$ all bubble spacetimes are identical, with
$a(\tau)=\tau$.) Of course, the FRW bubble universes are infinite, and
we could take the limit $R\to\infty$. But the constant factor $R^3$
drops out of the relative probabilities, so the value we choose for
the length scale $R$ is unimportant.

In all bio-friendly bubbles, there should be a period of internal
inflation, characterized by a large expansion factor $Z_j\gg 1$. After
the vacuum energy is thermalized, a certain number ${\cal N}_j^*$ of
observers will evolve per unit thermalized volume; its value will
depend on the local parameters of the low-energy physics. Counting
only regular observers, as it was done in \cite{GSPVW}, we can write
\beq
f_j\sim Z_j^3 {\cal N}_j^*.
\label{fN}
\eeq
But now we know that if the vacuum energy is positive inside the
bubble, then, apart from the regular observers, there are freak
observers who nucleate at a constant rate per unit spacetime
volume. This means that, if the freaks are included, then
$n_j(\tau)\to const$ and Eq.~(\ref{fj}) gives $f_j\to\infty$.
Moreover, as already mentioned in the Introduction, bubbles of false
vacuum will also nucleate at a constant rate, each bubble contributing
infinite numbers of both the regulars and the freaks.

Clearly, the prescription (\ref{fj}) is not acceptable as it
stands. The intent of the original formulation in \cite{GSPVW} was to
count only regular observers who evolve in the wake of bubble
nucleation. But the question is: On what basis can we discriminate
against the freak observers? It is not enough to say that they are
formed by quantum fluctuations. In models of cosmic inflation,
galaxies and other cosmic structures owe their eistence to quantum
fluctuations, so human observers may share the fuzzy quantum origin
with the freaks. In the next Section I will suggest a possible way of
regulating the infinity in Eq.~(\ref{fj}).

\section{Getting rid of freak observers}

My proposal is that there should be a sharp division between the kinds
of objects counted in $p_j$ and those counted in $f_j$. $p_j$ counts
the objects like bubbles, which nucleate in a vacuum at a constant
rate. These are equilibrium vacuum fluctuations in de Sitter
space. $f_j$ counts observers that arise due to non-equilibrium processes in
the wake of a quantum fluctuation of type $j$. This can still be
expressed by Eq.~(\ref{fj}) if we make the replacement
\beq
n_j(\tau)\to {\tilde n}_j(\tau)=n_j(\tau)-n_j^{(eq)}.
\label{neq}
\eeq
Here, $n_j(\tau)$ is the total production rate of observers,
$n_j^{(eq)}=const$ is the equilibrium rate at which they are produced
by quantum fluctuations in a de Sitter vacuum, and the difference
${\tilde n}_j(\tau)$ is the production rate due to non-equilibrium
processes.  With this replacement, the integral in (\ref{fj}) is
convergent and can still be estimated by Eq.~(\ref{fN}).

To put it slightly differently, the events counted in $p_j$ are {\it
uncaused}, random, equilibrium fluctuations. The factor $f_j$ assigns
a weight to these fluctuations, based on the average number of
observers formed as a result of non-equilibrium processes {\it caused}
by a fluctuation of type $j$. A fluctuation producing one isolated
freak observer gets a weight of 1. Even if there is a huge fluctuation
producing a large number of freaks, the weight will always be
finite. On the other hand, a bio-friendly bubble produces an infinite
number of observers and thus has an infinite weight. Formally, this
can be accounted for by taking the limit $R\to\infty$ in
Eq.~(\ref{fj}). As a result, freak observers get a vanishing relative
weight, while the relative weights of the bubbles are independent of
$R$.

\section{Discussion}
 
The purpose of this note is to clarify the pocket-based measure
(\ref{Pj}) of Ref.~\cite{GSPVW}. My proposal is that the 
factor $p_j$ should be interpreted as the abundance of equilibrium
vacuum fluctuations of a given type. The selection weight $f_j$ is
proportional to the average number of observers formed due to
non-equilibrium processes resulting from such fluctuations. Freak
observers are produced either individually or in finite groups, while
each bio-friendly bubble produces an infinite number of
observers. Thus, freak observers have a vanishing relative weight and
do not contribute to the measure, even though their nucleation rate
may in some cases be higher than that of the bubbles.

I should finally mention some open issues. The pocket-based measure,
as it is presently formulated, assumes that bubbles do not collide
with one another. 
Also, the current formulation cannot be directly applied to models
where pockets are formed by quantum diffusion. Some ideas toward
extension to this class of models have been discussed in
\cite{GSPVW}. 

The pocket-based measure does account for bubble formation within
bubbles. However, it is tacitly assumed that secondary bubbles
($s$-bubbles) do not interfere with the evolution of observers in the
primary bubble ($p$-bubble). This should be a good approximation if
the $s$-bubble formation rate is very low. The production rate of
observers in Eq.~({\ref{neq}) approaches zero when the stars die out
and other non-equilibrium processes in the $p$-bubble come to a
halt. We assume that $s$-bubbles that nucleate during the period
when ${\tilde n}_{p}(\tau)$ is substantially different from zero
affect only a small fraction of volume in the open FRW universe of the
$p$-bubble. Inclusion of this effect should result in a slight
renormalization of the selection factor $f_p$.

The formation rate of freak observers is likely to be enhanced for
some period of time $\tau$ inside the bubbles, either due to thermal
fluctuations (while the temperature is still higher than $T_{GH}$) or
to quantum fluctuations induced by time-varying fields and other
non-equilibrium processes in the wake of bubble nucleation. This is
good news for the freaks: their formation rate, given by
Eq.~(\ref{neq}), is non-zero after all. However, just as in the case
of regular observers, this rate approaches zero at large $\tau$, and
the fraction of freaks relative to the regular observers is likely to
be very small. The nucleation rate of $s$-bubbles may also be enhanced
at early times. This will probably result in some additional
renormalization of $p_s$ and $f_p$. This issue requires further study.

\acknowledgments

My views on these matters have been shaped in numerous discussions
with Jaume Garriga. I am also grateful to Raphael Bousso, Ben Freivogel and
Roni Harkin for stimulating discussions during my visit to Berkeley. 
This work was supported in part by grant RFP1-06-028 from The
Foundational Questions Institute.


\begin{thebibliography}{99}
\bibitem{Rees} 
To my knowledge, Martin Rees \cite{Rees1} was the first to discuss
disembodied brains arising from thermal or quantum fluctuations, in
connection with the Boltzmann's idea that we are observing a huge
fluctuation in the otherwise thermalized universe. The term
``Boltzmann's brains'' was later coined by Albrecht and Sorbo
\cite{Albrecht}, who also discussed the issue in relation to
inflationary cosmology.
\end{thebibliography}
\end{document}